\documentclass[12pt]{article}
\usepackage{citesort,fullpage,epsfig,psfrag,graphics,amsbsy,amssymb,amsmath}
\usepackage{caption}
\newcommand{\beq}{\begin{equation}}
\newcommand{\eeq}{\end{equation}}
\newcommand{\bea}{\begin{eqnarray}}
\newcommand{\eea}{\end{eqnarray}}
\newcommand{\bfs}{\boldsymbol}

\newcommand{\Tr}{{\rm Tr}}
\newcommand{\be}{\begin{equation}}
\newcommand{\ee}{\end{equation}}
\newcommand{\bq}{\begin{eqnarray}}
\newcommand{\eq}{\end{eqnarray}}
\newcommand{\ket}[1]{|#1\rangle}

\def\math{\mathsurround=0pt }
\def\leftrightarrowfill{$\math \mathord\leftarrow \mkern-6mu 
 \cleaders\hbox{$\mkern-2mu \mathord- \mkern-2mu$}\hfill
 \mkern-6mu \mathord\rightarrow$}
\def\overleftrightarrow#1{\vbox{\ialign{##\crcr
     \leftrightarrowfill\crcr\noalign{\kern-1pt\nointerlineskip}
     $\hfil\displaystyle{#1}\hfil$\crcr}}}

\let\l=\lambda

 \def\bd{\begin{document}} \def\ed{\end{document}}
\def\ds{\documentstyle} \let\fr=\frac \let\bl=\bigl \let\br=\bigr
\let\Br=\Bigr \let\Bl=\Bigl
\let\bm=\bibitem
\let\na=\nabla
\let\pa=\partial \let\ov=\overline
\def\ft#1#2{{\textstyle{{\scriptstyle #1}\over {\scriptstyle #2}}}}
\def\fft#1#2{{#1 \over #2}}
\def\vp{\varphi}
\def\sst#1{{\scriptscriptstyle #1}}
\def\oneone{\rlap 1\mkern4mu{\rm l}}
\def\td{\tilde}
\def\wtd{\widetilde}
\def\dalemb#1#2{{\vbox{\hrule height .#2pt
        \hbox{\vrule width.#2pt height#1pt \kern#1pt
                \vrule width.#2pt}
        \hrule height.#2pt}}}
\def\square{\mathord{\dalemb{6.8}{7}\hbox{\hskip1pt}}}
\def\wtd{\widetilde}
\def\R{\rlap{\rm I}\mkern3mu{\rm R}}
\def\im{{\rm i}}
\def\tilg{\tilde{g}}
\def\tilF{\tilde{F}}
\def\tilA{\tilde{A}}
\def\varf{\varphi}
\def\tilf{\tilde{\phi}}
\def\tilh{\tilde{h}}
\def\rme{{\rm e}}
\def\ep{\epsilon}
\def\0{{(0)}}
\def\9{{(9)}}
\def\8{{(8)}}
\def\7{{(7)}}
\def\6{{(6)}}
\def\5{{(5)}}
\def\4{{(4)}}
\def\3{{(3)}}
\def\2{{(2)}}
\def\1{{(1)}}
\newcommand{\trace}{{\rm Tr}}
\newcommand{\ub}{\overline{U}}
\newcommand{\vb}{\overline{V}}
\newcommand{\uh}{\widehat{U}}
\newcommand{\vh}{\widehat{V}}
\newcommand{\ubh}{\overline{\widehat{U}}}
\newcommand{\vbh}{\overline{\widehat{V}}}
\newcommand{\lb}{\bar{\l}}
\newcommand{\Fb}{\overline{F}}
\newcommand{\Fh}{\widehat{F}}
\newcommand{\Fbh}{\overline{\widehat{F}}}
\newcommand{\Ab}{\overline{A}}
\newcommand{\Ah}{\widehat{A}}
\newcommand{\Abh}{\overline{\widehat{A}}}
\newcommand{\Gb}{\overline{G}}
\newcommand{\Gh}{\widehat{G}}
\newcommand{\Gbh}{\overline{\widehat{G}}}
\newcommand{\Pb}{\overline{P}}
\newcommand{\Ph}{\widehat{P}}
\newcommand{\Pbh}{\overline{\widehat{P}}}
\newcommand{\Qb}{\overline{Q}}
\newcommand{\Qh}{\widehat{Q}}
\newcommand{\Qbh}{\overline{\widehat{Q}}}
\newcommand{\Bb}{\overline{B}}
\newcommand{\Bh}{\widehat{B}}
\newcommand{\Bbh}{\overline{\widehat{B}}}
\newcommand{\fhns}{\hat{F}^{\rm (NS)}}
\newcommand{\fhrr}{\hat{F}^{\rm (RR)}}
\newcommand{\ahns}{\hat{A}^{\rm (NS)}}
\newcommand{\ahrr}{\hat{A}^{\rm (RR)}}
\newcommand{\hhrr}{\hat{H}^{\rm (RR)}}
\newcommand{\hchi}{\hat{\chi}}
\newcommand{\hphi}{\hat{\phi}}
\newcommand{\htau}{\hat{\tau}}
\newcommand{\cG}{{\cal G}}
\newcommand{\cGb}{\overline{{\cal G}}}
\newcommand{\cH}{{\cal H}}
\newcommand{\cP}{{\cal P}}
\newcommand{\cPb}{\overline{{\cal P}}}
\newcommand{\cQ}{{\cal Q}}
\newcommand{\cQb}{\overline{{\cal Q}}}
\newcommand{\cM}{{\cal M}}
\newcommand{\cN}{{\cal N}}
\newcommand{\cO}{{\cal O}}
\newcommand{\cD}{{\cal D}}
\newcommand{\cL}{{\cal L}}
\newcommand{\vpp}{\mbox{$\langle{\scriptstyle++}\rangle$}}
\newcommand{\vmp}{\mbox{$\langle{\scriptstyle-+}\rangle$}}
\newcommand{\vppp}{\mbox{$\langle{\scriptstyle+++}\rangle$}}
\newcommand{\vmpp}{\mbox{$\langle{\scriptstyle-++}\rangle$}}
\newcommand{\vpmp}{\mbox{$\langle{\scriptstyle+-+}\rangle$}}
\newcommand{\goesas}[1]{{}_{{\displaystyle\sim}\atop#1}}
\newcommand{\impliesas}[1]{{}_{{\displaystyle{\Longrightarrow}}\atop#1}}
\renewcommand{\thepage}{\arabic{page}}

\begin{document}
\setlength{\captionmargin}{36pt}
%\begin{titlepage}
\begin{titlepage}
\begin{flushright}
\phantom{UFIFT}%\timestamp
\end{flushright}

\vskip 3cm
\begin{center}
\begin{large}
{\bf Scientific Biography of Stanley Mandelstam, Part I: 1955-1980}
\end{large}
\vskip 2cm
{\large 
Charles B. Thorn
\footnote{E-mail  address: {\tt thorn@phys.ufl.edu}}
}
\vskip0.20cm
{\it Institute for Fundamental Theory,\\
Department of Physics, 
University of Florida
Gainesville FL 32611
}

\vskip24pt
\end{center}
\begin{abstract}
\noindent \end{abstract} 
I review Stanley Mandelstam's many contributions
to particle physics, quantum field theory and string theory covering the
years 1955 through 1980. His more recent work will be reviewed by
Nathan Berkovits. This is my contribution to the Memorial Volume for
Stanley Mandelstam (World Scientific, 2017).
\vfill
\end{titlepage}
\section{Introduction}
Stanley Mandelstam's career in theoretical physics spans nearly 60 years
of path-breaking research in the theory of elementary particles.
He was an undisputed master of quantum field theory (QFT), which, to this day,
is accepted as the indispensable framework for describing
relativistic quantum mechanics,

By the time Stanley began his research in the mid 1950's
\cite{Mandelstam:1955}, quantum 
electrodynamics (QED), the QFT of the electromagnetic
field and charged particles, was well-understood in the context
of perturbation theory. This success was based on the good fortune that
the expansion parameter $\alpha=e^2/(4\pi\hbar c)\approx1/137$ is
quite small. From the beginning, the problems Stanley set out to
illuminate involved aspects of QFT that were not amenable to
perturbation theory. In the early 50's it was evident that the
strong interactions (the physics of the nuclear force), 
if described by QFT, would involve an
expansion parameter (analogous to $\alpha$) which was not small.
Thus the strong interactions became an early focus of Stanley's
considerable talents.

His deep understanding of the analytic structure of the individual
terms of perturbation theory (Feynman diagrams) led him to a proposal for a
representation of the scattering amplitude for pions and nucleons,
now known as the Mandelstam representation \cite{Mandelstam:1958,Mandelstam:1959}. 
This proposal became
the mainstay of the $S$-matrix approach to the theory of strong
interactions espoused by Geoffrey Chew. Together with Geoff,
Stanley played a pivotal role in laying the foundations for this program
\cite{Mandelstam:1960a,Mandelstam:1960b,Mandelstam:1960c}

Although initially $S$-matrix theory focused on seeking
self-consistent  solutions based on Lorentz invariance, unitarity,
crossing symmetry and the analyticity of the $S$ matrix in the 
Lorentz scalars $p_k\cdot p_l$\footnote{Minkowski scalar products
 formed from the 4-momenta
$p_k$ of the scattered particles in the initial and final states},
the ideas of Tullio Regge on the analyticity of partial wave amplitudes
in complex angular momentum ($J$) quickly became a source of deeper
insights, especially with respect to the high energy behavior of
scattering amplitudes\cite{Mandelstam:1961a,Mandelstam:1963a,c1,c2,c3,Mandelstam:1963b,Mandelstam:1965a}.

Stanley's unmatched understanding of the analytic structure of
Feynman diagrams soon elucidated just how intricate the 
singularity structure of the complex $J$-plane was
\cite{Mandelstam:1963a,Mandelstam:1963b}. There were
cuts in $J$ which shielded the physics from the dire consequences
of the essential singularity found by Gribov and Pomeranchuk.
Furthermore, using his eponymous representation, he showed that
all of these complications required the presence of the
``third double spectral function'' $\rho_{tu}$. He then suggested a
new approximation scheme in which $\rho_{tu}$ is neglected in first instance
\cite{Mandelstam:1967b} and then brought back in perturbatively \footnote{Years later (1973) Gerard.'t Hooft \cite{thooftlargen}
identified the expansion parameter for non-planar
contributions as $1/N$ for an $SU(N)$ nonabelian gauge theory.} . 
This was a crucial insight.
The second and third double spectral functions are absent from 
the sun of all planar Feynman
diagrams. So a systematic approximation scheme could be formulated:
First sum all the planar diagrams to a given physical process,
and then bring in nonplanar diagrams in a perturbative expansion.
It was at least consistent with known facts to hypothesize that
the sum  of planar diagrams has only simple Regge poles in the
$J$-plane. Stanley proposed \cite{Mandelstam:1967a,Mandelstam:1968c} 
that one seek
an approximation in which Regge trajectories are linear
and, simultaneously, resonances are narrow. The latter means that the
only singularities of scattering amplitudes are
 poles in the momentum variables $p_k\cdot p_l$, located on the real axis. 

A``bootstrap'' solution
based on these principles was then the goal of many young theorists,
including Dolen, Horn, and Schmid from a phenomenological point
of view \cite{dhs}. 
Ademollo, Rubinstein, Veneziano, and Virasoro \cite{ademollorvv}
were
soon in hot pursuit of such a bootstrap. This activity culminated
in Veneziano's discovery \cite{veneziano} of his four point amplitude 
and the rest, as they say, is history. String theory was born and in the ensuing
decades has developed into a promising approach to quantum gravity.

Stanley embraced these exciting developments whole-heartedly
\cite{Mandelstam:1968e,Mandelstam:1969a,Mandelstam:1969c}.
He sketched how generalizations of Veneziano's formula
could be used to build a relativistic quark model of mesons
and baryons \cite{Mandelstam:1969e,Mandelstam:1970a,Mandelstam:1970b,Mandelstam:1970c}. He built on the lightcone quantization 
of the Nambu-Goto string \cite{goddardgrt} to show that 
the dual resonance models
were indeed the scattering amplitudes of strings \cite{mandelstambosonic}. 
He used his
lightcone interacting string formalism to accomplish a
theoretical tour-de-force: the calculation of scattering
amplitudes involving four or more fermions 
\cite{mandelstamnsr,Mandelstam:1985c}, and then to complete
the calculation of all multi-loop diagrams in string theory
\cite{mandelstamdet,Mandelstam:1986a,Mandelstam:1991a}.

These were all valiant efforts in aiding string theory {\it per se}.
But Stanley never dropped his fascination for the strong interactions,
which had indirectly led to all these theoretical developments. 
By the early 1970's,
with the discovery of asymptotic freedom and charm, 
it became abundantly clear that
the strong interactions should be described by a non-abelian
gauge theory of quarks and gluons (QCD). The catch was that quarks and gluons
should be permanently trapped inside of baryons and mesons--
the hypothesis of quark confinement. So it was natural
for Stanley to attack the problem of quark confinement. Motivated by
an analogy with a superconductor's confinement of magnetic monopoles
due to the condensation of a charged field, he developed a
``dual'' scheme for the condensation of color-magnetic monopoles
in QCD \cite{Mandelstam:1974c,Mandelstam:1974d,Mandelstam:1978a,Mandelstam:1978b,Mandelstam:1979a,Mandelstam:1979b,Mandelstam:1980a,Mandelstam:1982a}

But still more achievements were in store. As a spinoff from his work
on confinement, Stanley 
illuminated the equivalence
of the 2 dimensional massive Thirring model to the 2 dimensional
Sine-Gordon QFT by constructing an explicit operator mapping between the
two QFT's \cite{Mandelstam:1975a,Mandelstam:1976a}. This nice piece of mathematical physics provided a toy
prototype of the duality between magnetic and electric confinement
in nonabelian gauge theories.
Then supersymmetry became a hot topic in the late 1970's and early 1980's. 
There were exciting conjectures, such as
that ${\cal N}=4$ supersymmetric non-abelian gauge theory should be
finite in the ultraviolet. Naturally, Stanley was one of the first to 
construct a proof of this deep conjecture. 

In sections 2 through 7 below 
I endeavor to explain Stanley's important
achievements through the 1970's. The final section 8 is
my personal tribute to him emphasizing his role 
as one of the pioneers of string theory.
My colleague Nathan Berkovits will write Part 2 of his scientific biography,
describing Mandelstam's work since 1980.
\section{Double Dispersion Relations}
A general scattering process is a reaction
\bea
P_1+P_2\longrightarrow P_3+P_4+\cdots P_N
\eea
where each letter $P_k$ stands for a different particle participating in the
process either in the initial or final state. It is
described by an amplitude ${\cal A}$ which is a function of
the Lorentz scalars $p_i\cdot p_j$ formed from
the energies and momenta of all the particles. Here $-p_i^2=m_i^2$
where $m_i$ is the mass of the $i$th particle which is not variable, so 
we can, without loss of information, take $i<j$. In addition,
energy and momentum is conserved so that there are only $N-1$ independent
$p_i$. If we eliminate $p_N=-p_1-\cdots-p_{N-1}$ then the mass shell
constraint $p_N^2=-m_N^2$ can be rewritten
\bea
-\sum_{i<j<N}(p_i+p_j)^2=\sum_{i=1}^Nm_i^2
\eea
Mandelstam focused on the case $N=4$ and defined the three invariant
variables
\bea
s&=&-(p_1+p_2)^2,\qquad t=-(p_2+p_3)^2,\qquad u=-(p_1+p_3)^2\\
s+t+u&=& m_1^2+m_2^2+m_3^2+m_4^2
\eea
any two of which can be chosen as independent variables. 

\begin{figure}[ht]
\begin{center}
\includegraphics[width=4in]{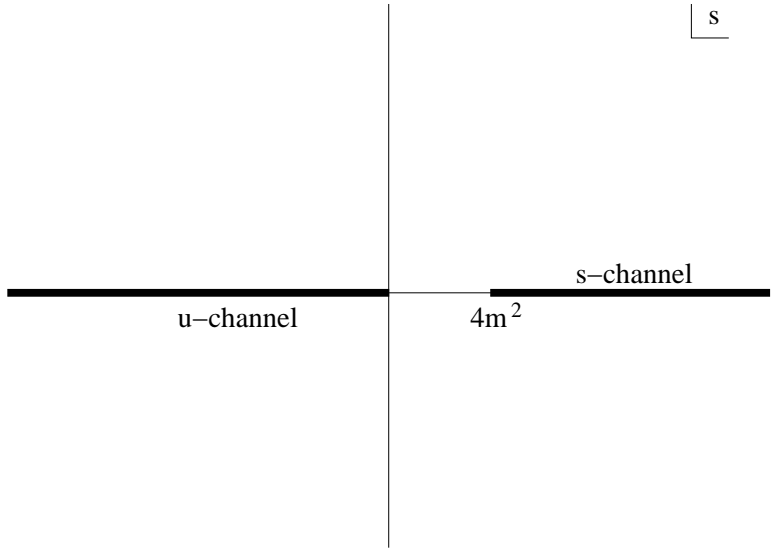}
\caption{The complex $s$ plane at $t=0$. The thick lines indicate branch cuts.
the right branch point is due to the threshold at $s=4m^2$, the left branch
point is
due to threshold at $u=4m^2=4m^2-s-t$, and is shown for $t=0$.}
\label{splane}\end{center}
\end{figure}

For simplicity let's assume that all masses are the same, so that $s+t+u=4m^2$. 
Then crossing symmetry is the statement that ${\cal A}$ describes three
possible scattering processes, depending on the range of $s,t,u$
The $s$-channel is $P_1+P_2\to P_3+P_4$, which requires $s>4m^2$
and $t,u<0$; the $t$-channel is $P_1+{\bar P}_{4}\to {\bar P}_2+P_{3}$,
which requires $t>4m^2$ and $s,u<0$; and the $u$ channel is
$P_1+{\bar P}_{3}\to {\bar P}_2+P_{\bar4}$ which requires $u>4m^2$ and $s,t<0$.
In this reaction notation the bar denotes the antiparticle.
Since the three channels are in nonoverlapping regions of $s,t,u$,
applying crossing symmetry requires continuing somehow between these
regions. In S-matrix theory one postulates that ${\cal A}$ is an analytic
function of $s,t,u$ in a domain of the complex planes which contains all three
regions. For example at $t=0$ the postulated domain of analyticity in
the $s$ plane is shown in Fig.\ref{splane}.
A single dispersion relation is obtained from the analyticity hypothesis
by first writing the Cauchy formula
\bea
{\cal A}(s)=\frac{1}{2\pi i}\oint_C ds^\prime\frac{{\cal A}(s^\prime)}{s^\prime -s}
\eea
where $s$ is in the domain of analyticity, and the contour $C$ is an
infinitesimal circle centered on $s$. Then one deforms $C$ to large semicircles
at infinity plus straight lines just above and just below the cuts. The result is a formula like
\bea
{\cal A}(s)=\frac{1}{2\pi i}\int_{4m^2}^\infty ds^\prime \frac{{\cal D}_R}{s^\prime -s}+\frac{1}{2\pi i}\int_{-\infty}^0ds^\prime \frac{{\cal D}_L}{s^\prime -s}
+ {\rm semicircle~terms}\eea
where ${\cal D}_{R,L}$ are the discontinuities across the right and
left cuts. Such relations
have experimental implications. For example for $t=0$ the discontinuities can
be related by the optical theorem to total cross sections in the respective
scattering channels and the semicircle terms will go to zero if 
${\cal A}$ vanishes sufficiently rapidly at infinity. The dispersion relation
then expresses  ${\cal A}(s,0)$ in terms of an integral over
the total cross section.

\begin{figure}[ht]
\begin{center}
\includegraphics[width=4in]{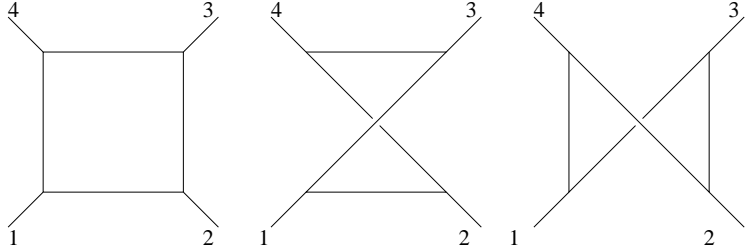}
\caption{The three single loop box Feynman diagrams, which are the simplest
possessing the nonzero double spectral functions, $\rho_{st}$, $\rho_{su}$,
or $\rho_{tu}$ respectively. Relative to the $s$ channel, the last one is the
``third'' double spectral function. Note that since the cyclic order of
the particle labels is the same in all three diagrams, 
the second and third diagrams have crossed
lines.}
\label{boxes}
\end{center}
\end{figure}

Mandelstam's seminal contribution \cite{Mandelstam:1958,Mandelstam:1959} 
was to exploit the postulate of simultaneous analyticity
in more than one variable--two in the case of $N=4$ to obtain
double dispersion relations. The right sides are now double integrals,
and the Mandelstam representation for equal masses takes the form
\bea
{\cal A}&=&\frac{1}{\pi^2}\int_{4m^2}^\infty ds^\prime dt^\prime
\frac{\rho_{st}(s^\prime,t^\prime)}{(s^\prime-s)(t^\prime-t)}
+\frac{1}{\pi^2}\int_{4m^2}^\infty ds^\prime dt^\prime
\frac{\rho_{su}(s^\prime, u^\prime)}{(s^\prime-s)(u^\prime-u)}\nonumber\\
&&\qquad+\frac{1}{\pi^2}\int_{4m^2}^\infty ds^\prime dt^\prime
\frac{\rho_{tu}(t^\prime,u^\prime)}{(t^\prime-t)(u^\prime-u)}
\eea
Here we have assumed sufficient fall off at infinity and have not
included pole terms. The analyticity assumed here is not universally true:
there are situations where singularities occur at complex values
of $s,t,u$ for which the integrals would not be over real variables. But
it does reflect the analyticity of low orders in perturbation theory,
for example that of the box diagrams in Fig.\ref{boxes}
The Mandelstam representation explicitly exhibits the singularities in $s,t,u$
that are expected on the basis of particle thresholds and experience
with Feynman diagrams. It therefore proved an indispensable tool in
the development of Chew's program of $S$-matrix theory
\cite{Mandelstam:1960a,Mandelstam:1960b,Mandelstam:1960c}. 
\section{Complex Angular Momentum}The striking
way the Mandelstam representation separates 
the singularities of the four point amplitude into
three distinct terms, characterized by the double spectral
functions $\rho_{st}$, $\rho_{su}$, and $\rho_{tu}$, has also played
a central role in understanding the singularity
structure of partial wave amplitudes continued into the complex angular momentum
plane, as envisioned by Tullio Regge.

Partial wave amplitudes trade dependence on angle for dependence on angular momentum. In scattering by a potential the scattering amplitude $f(E,\theta)$
can be expanded in Legendre polynomials thus
\bea
f(E,\theta)=\sum_{J=0}^\infty a_J(E)P_J(\cos\theta)
\eea
where $a_J$ are the partial wave amplitudes, which completely characterize
the scattering. When $J$ is an integer, $a_J(E)$ has a pole at the energy
$E_{J,n}$ of each bound state with angular momentum $J$:
\bea
a_J&\sim& \frac{R_{Jn}}{E-E_{Jn}},\qquad E\sim E_{Jn}
\eea
When $a_J$ is continued into the complex $J$-plane, $E_{Jn}$
becomes an analytic function of $J$, and the pole location moves.
Viewed as a pole in the $J$ plane, it becomes a Regge pole $\alpha(E)$
whose location depends on energy. The contribution of a Regge
pole to the scattering amplitude is proportional to
$P_{\alpha(E)}(\cos\theta)$ which has the large $\cos\theta$ behavior
$(\cos\theta)^{\alpha(E)}$, known as Regge asymptotic behavior.

\begin{figure}[ht]
\begin{center}
\includegraphics[width=4in]{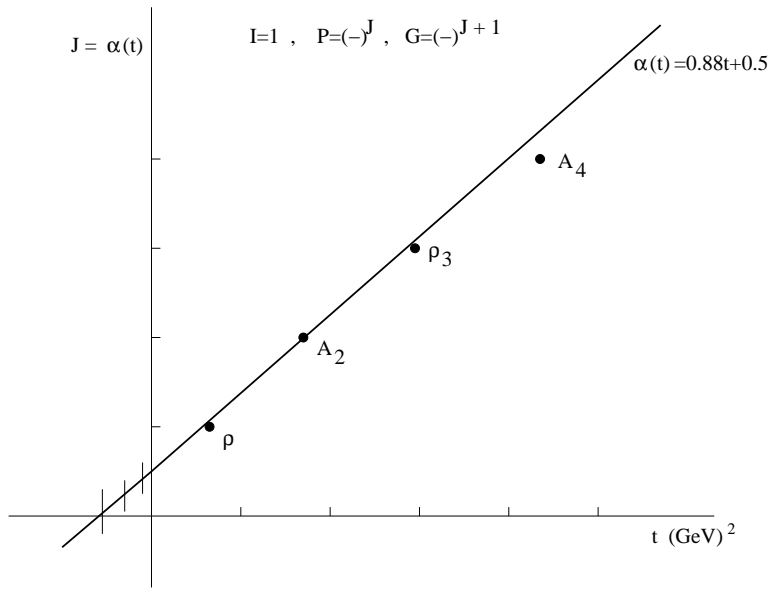}
\caption{Chew-Frautschi plot for isospin 1 exchange in the $t$-channel. 
The only data points for $t>0$  
are the angular momentum of particles of mass $\sqrt{t}$ 
with the quantum numbers ($I=1$) 
of the $t$-channel. In the early
sixties the only particle on the plot was the $\rho$ meson.
For $t<0$, $t$ is in the physical region of $s$ channel scattering, 
and $\alpha(t)$
can be inferred from experiments at large positive $s$. The error
bars are only to roughly indicate the measurement uncertainties. 
The experimental near
linearity of Regge trajectories, especially for those with non-vacuum
quantum numbers was very suggestive.}
\label{chewf}
\end{center}
\end{figure}

Chew, Frautschi, and Mandelstam \cite{Mandelstam:1961a} realized the importance of these
ideas in relativistic $S$-matrix theory. The momentum transfer invariant
$t=(s-4m^2)(\cos\theta-1)/2$, so large $\cos\theta$ is identified
with large $t$. The invariant $s$ is related to the scattering
energy, so in a relativistic context, Regge behavior is interpreted
as
\bea
A(s,t)&\sim \beta(s)(t)^{\alpha(s)},\qquad {\rm as} \quad t\to\infty
\eea
In $t$ channel scattering this links high energy (large $t$) 
to the exchange of the Regge trajectory $J=\alpha(s)$ in the
momentum transfer variable ($s<0$ in the $t$-channel).
This was not only a breakthrough in theory, but it also had
direct experimental implications. One could extract
the trajectories $\alpha(s)$ for $s<0$ by careful measurement of high
energy (large $t$) cross sections at varying momentum transfer $s$. 
The trajectory
$\alpha(s)$ is an extrapolation of the relation between angular momentum
and mass squared of the particles at discrete positive values of
$s$. These can be dramatically presented on a Chew-Frautschi plot
as shown in the Fig.\ref{chewf} for which we have switched the
roles of $s$ and $t$.
While physical angular momentum is necessarily nonnegative,
the extrapolation $\alpha(t)$ to negative $t$ could well become negative, 
with welcome implications for the convergence of dispersion relations.

\begin{figure}[ht]
\begin{center}
\includegraphics[width=6in]{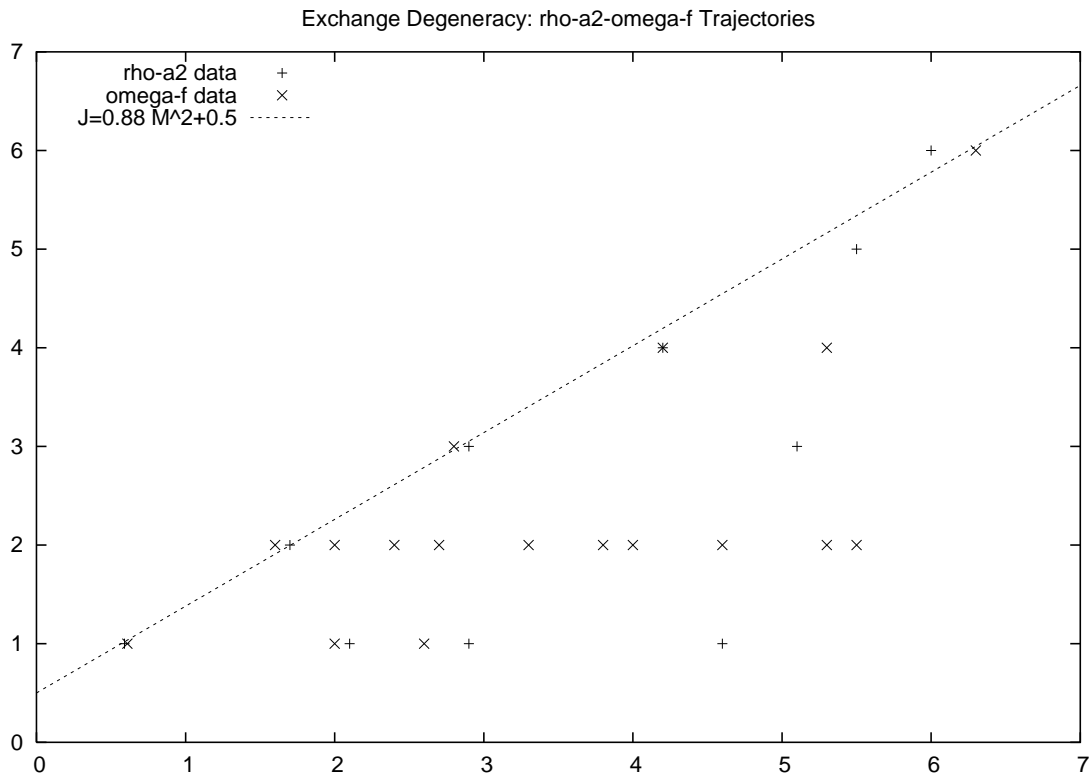}
\caption{Evidence for exchange degeneracy. We plot angular momentum
versus mass squared for known resonances with isospin 1 and 0
and both even and odd angular momentum. The ones with highest $J$
fall very close to the same linear Regge trajectory. Exchange degeneracy
would be a consequence of the dominance of a single double spectral
function, or, alternatively, of the dynamics of planar Feynman diagrams.}
\label{exchangedeg}
\end{center}
\end{figure}

A subtlety of the continuation to complex angular momentum in the
relativistic case is that even and odd angular momentum must,
in principle, be continued separately. This is due to the presence
of both $st$ and $su$ double spectral functions. As a consequence
Regge trajectories in the even signature (odd signature) 
continuation only imply particles
at even (odd) values of $J$. There is no {\it a priori} relation between the
even and odd trajectories. However, the Chew-Frautschi plots for
various trajectories (see Fig.\ref{exchangedeg})
showed near coincidence of even and odd
trajectories. This exchange degeneracy suggested that some processes
were dominated by only one double spectral function.

Although the promising phenomenological success of Regge pole phenomenology
generated much excitement, it soon became clear that the complex $J$ plane
has a much more complicated singularity structure than simply Regge poles.
In spring 1962 Amati, Fubini, and Stranghellini (AFS) argued for the presence
of Regge cuts in processes involving the exchange of two Regge poles.
Later in the same year
Gribov and Pomeranchuk (GP) uncovered the apparent need for an essential
singularity in the complex $J$ plane, in the form of an accumulation
point of Regge poles, near $J=-1$ for scattering of scalar
(spin 0) particles. In \cite{Mandelstam:1963a} Mandelstam
showed that this phenomenon occurs at nonnegative $J=2s-1$ 
for processes involving particles with spin $s$. 
Besides the difficulties these complications
caused for the experimental interpretation of Regge poles, they also raised
serious theoretical questions. 

\begin{figure}[ht]
\begin{center}
\includegraphics[width=4in]{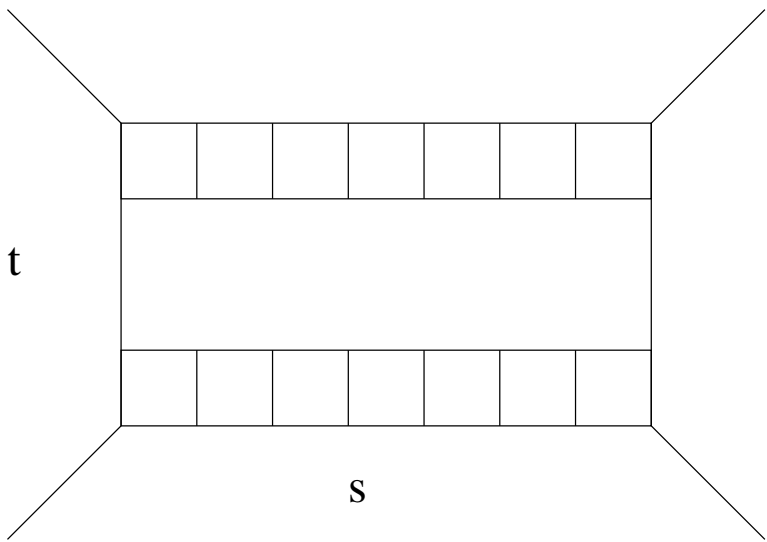}
\caption{Feynman diagrams reflecting the AFS argument for Regge cuts.
the ladder structures are meant to be summed over all numbers of rungs
so that they represent a composite structure describing the exchange of
a Regge pole. AFS argued that the exchange of two Regge poles would produce a Regge cut. Mandelstam showed that the apparent cut in these diagrams
is not present on the physical sheet due to the planarity of the diagrams.}
\label{planarcuts}
\end{center}
\end{figure}

\begin{figure}[ht]
\begin{center}
\includegraphics[width=4in]{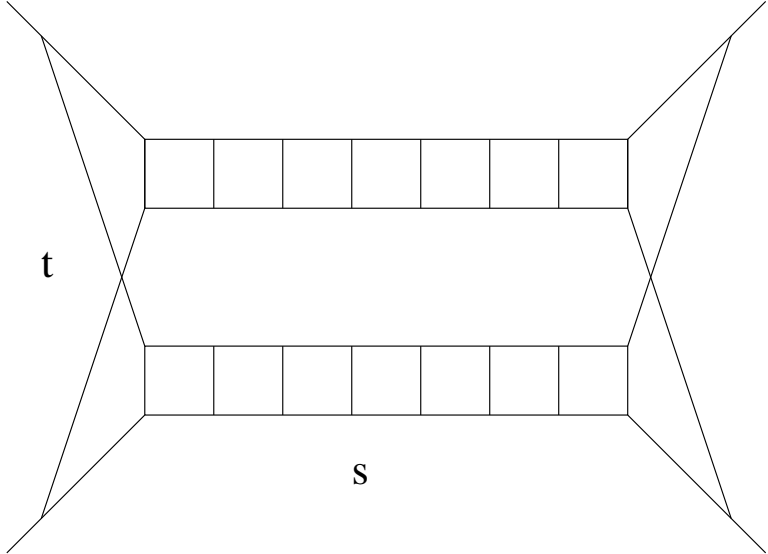}
\caption{Diagrams for which a Regge cut is present on the physical sheet.
The nonplanar structures on either side of the diagram reflect the
contribution of the third double spectral function $\rho_{su}$
in the $t$-channel. If one perturbs in nonplanarity, e.g. in powers of the
$\rho_{su}$, Regge cuts first appear at second order $O(\rho_{su}^2)$.}
\label{nonplanarcuts}
\end{center}
\end{figure}

Mandelstam, in a tour-de-force, clarified these issues in a way that
indicated how one might be able to gain control over the
ensuing complications \cite{Mandelstam:1963b}. He first showed that
the Feynman diagrams considered by AFS did not in fact possess the 
angular momentum cuts they had argued for. Those diagrams involved the
exchange of two ladder sub-diagrams as shown in Fig.\ref{planarcuts} 
And because they were planar the potential cuts necessarily cancelled out
on the physical sheet, and, moreover weren't afflicted with the
GP phenomenon. He identified the key reason, namely
that planar diagrams lacked third double spectral functions.
He then showed that with sufficient nonplanarity as in Fig.\ref{nonplanarcuts}
the cuts remained on the physical sheet. Finally he showed that
the cuts of this type screened  the physical sheet from
the essential singularities discovered by GP.
These two papers definitively resolved the theoretical issues posed by
AFS and GP. 

Moreover, he pinned down how these complications of 
complex angular momentum arise: they appear only in
amplitudes possessing third double spectral functions. If
such contributions to a given process were relatively small,
Regge pole phenomenology might be viable. In a later paper 
with Chau (formerly Wang) \cite{Mandelstam:1967b} they clarified further
the Gribov-Pomeranchuk phenomenon making use of precisely
this strategy--perturbing in the effects of the third double
spectral function. Feynman diagrams which involve such
effects have some degree of nonplanarity, as is evident 
from Figs.\ref{planarcuts} and \ref{nonplanarcuts}, 
suggesting that one way to perturb in them is
to take as first approximation the sum of only planar Feynman diagrams.

The most promising experimental processes for 
neglect of nonplanar effects involve nonzero
quantum number exchange, a famous example of which is the
charge exchange reaction $\pi^- p\to \pi^0 n$. An
insightful analysis of this process was reported in the summer
of 1967 by Dolen, Horn, and   
Schmid (DHS) \cite{dhs} on the assumption that Regge poles control the
high energy behavior.  The success of this analysis gave impetus 
to the work that led to dual resonance models (DRM).
\section{Narrow-Resonance/Regge Bootstrap}
In a seminal paper \cite{Mandelstam:1968c}, announced earlier at a conference
\cite{Mandelstam:1967a}, 
Mandelstam laid the
theoretical foundation which paved the way toward DRM.
He proposed that the bootstrap program in $S$-matrix theory
be implemented in stages. Motivated by the near linearity of
the Regge trajectories from analyses such as DHS, he noted that
exactly linear trajectories would imply that resonances lying
on the trajectory would be narrow (i.e. long-lived). This was because
the trajectory functions should satisfy  a dispersion relation
of the form
\bea
\alpha(s)=a s+b +\frac{1}{\pi}\int ds^\prime\frac{{\rm Im\ }\alpha}{s^\prime-s}
\eea
Exact linearity would then require ${\rm Im\ }\alpha=0$, and since the resonance
widths are proportional to ${\rm Im\ }\alpha$, they would have to be.
zero as well. Of course such an approximation means that
unitarity is only approximate. Assuming that third double spectral effects like 
Regge cuts and the GP phenomenon were absent and also that all
particles lie on Regge trajectories, the analyticity of the
$S$-matrix could be implemented in two ways. Focusing on an amplitude with
singularities in the $s>0$ and $t<0$ channels, one could either write the
amplitude as a sum of zero width resonance poles {\it or} write the
same amplitude
as a sum of $t$-channel Regge poles. These two expansions should agree:
which gives the approximate bootstrap equation
\bea
\sum_k\frac{R_k(t)}{s-m_k^2}&=&\sum_l \beta_l(t)s^{\alpha_l(t)}
\eea
This equality (called {\it duality} at the time)
should then lead to restrictions on the Regge parameters
and the particle masses. He went further by assuming only a finite
number of terms (typically 1 or 2) on each side leading to some
reasonable relations. The DHS analysis had confronted the data 
of pi nucleon charge exchange with either one (or two) resonances or
one (or two) Regge trajectories, and showed that either described the data well.
But the path-breaking proposal of Mandelstam's 
paper was that it should be possible 
to devise a narrow resonance approximation to Regge-behaved
scattering amplitudes provided that the Regge trajectories were
exactly linear. Relaxing exact unitarity made this bootstrap problem
much more tractable than the original one, which insisted on exact unitarity.
Indeed roughly a year later Veneziano discovered an exact solution
of Mandelstam's narrow resonance bootstrap: the celebrated Veneziano formula:
\bea
{\cal A}&=&g^2\frac{\Gamma(-\alpha(s))\Gamma(-\alpha(t))}{\Gamma(-\alpha(s)
-\alpha(t))},\qquad \alpha(x)=\alpha^\prime x+\alpha_0.
\eea
This formula is manifestly crossing symmetric under $s\leftrightarrow t$,
has only poles in $s$ or $t$ at $\alpha(s)=-n$, $n=0,1, \dots$,
and, using Stirling's formula can be shown to have Regge behavior
${\cal A}\sim \Gamma(-\alpha(t))(-s)^{\alpha(t)}$ as $s\to-\infty$
at fixed $t$.

Over the following year Mandelstam employed the Veneziano formula
in his narrow resonance Regge bootstrap program sketched above. One
of the salient features of the formula is the absence of $u$ channel poles:
in effect it possesses only the double spectral function $\rho_{st}$. 
This implies a planar structure in whatever underlying dynamics might
produce it. Meson spectroscopy is well explained by thinking of mesons as
quark-antiquark composites. To incorporate this quark structure in
his bootstrap, Mandelstam imagined the planar structure to be
bounded by a quark line, which should inject spin and flavor quantum
numbers. In this language the Veneziano model itself would correspond
to spinless flavor singlet quarks. It should then be supplemented with
spin and flavor factors corresponding to spin 1/2 quarks carrying
flavor SU(3) quantum numbers. In \cite{Mandelstam:1969e} Mandelstam
determined these factors by imposing crossing symmetry and factorization in
both $t$ and $s$ channels. As explained in the paper, an inevitable defect
of such a construction, due to the requirement of Lorentz invariance,
is a doubling of trajectories half of which are ghosts, meaning that
the particle pole residues due to them have the wrong sign.
Not yet realizing that ghost-free narrow resonance models would
eventually be found, Mandelstam speculated that the narrow
resonance approximation was responsible for such defects. In any
case the relativistic quark bootstrap was compared to experiment
under the assumption that these ghost trajectories wouldn't play a
significant role, with qualitative success.

According to the quark model, baryons are composites of three quarks,
so that baryonic processes can not be described by purely planar
models. In \cite{Mandelstam:1970a} Mandelstam constructed
what he called minimal nonplanar dual resonance models which
could describe the three quark structure of baryonic amplitudes.
he then used them to extend his relativistic quark model bootstrap 
to include such processes \cite{Mandelstam:1970b,Mandelstam:1970c}.
The minimal nonplanar dual models constructed here have received
little attention in the literature, and their further study might
well lead to interesting insights.

\section{Dual Resonance Models}
Veneziano's discovery precipitated an explosion of research activity directed
toward discovering and developing the underlying theory of 
the Veneziano amplitude.
Many researchers participated in the development of generalizations.
Mandelstam embraced, guided,  and participated in these rapid developments. 
Early on, he explored possible modifications of the original 4 particle
Veneziano amplitude \cite{Mandelstam:1968e}. He also clarified 
Virasoro's new four particle amplitude by giving it a
two dimensional integral representation analogous to a one-dimensional
integral representation of Veneziano's amplitude \cite{Mandelstam:1969a}.
This development presaged the eventual closed string interpretation of the
Virasoro model.

\begin{figure}[ht]
\begin{center}
\includegraphics[width=2in]{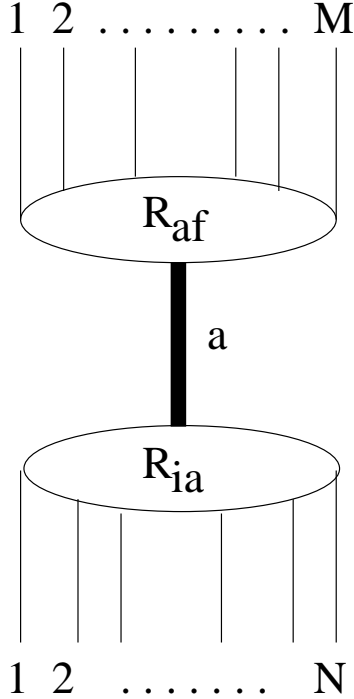}
\caption{Unitarity in the narrow resonance limit dictates that a resonance
pole in a process of $N$ particles to $M$ particles factor:
%\bea
${\cal A}\sim\sum_a{R_{ia}R_{af}}/{(s-m_R^2)}$
%\eea
where $m_R$ is the mass of the resonance, $R_{ia}$ is the scattering amplitude
for the process $i\to a$ and $R_{af}$ is the scattering amplitude for 
the process $a\to f$. In the
case of forward elastic scattering ($f=i$) $R_{ai}=R_{ia}^*$ 
and each factored term must be positive (No ghosts).}
\label{factor}
\end{center}
\end{figure}

Of course, with his early appreciation of the importance
of analyticity in multiple invariants describing a process, Mandelstam 
immediately realized that there should be generalizations to
narrow resonance amplitudes involving any number of particles.
The physical requirement of unitarity played an important guiding
role in these generalizations.
The narrow resonance approximation certainly abandons exact unitarity,
but scattering amplitudes
are still subject to unitarity constraints. The approximation
amounts to the assertion that scattering amplitudes are meromorphic
in the invariants $p_k\cdot p_l$, that is: the only singularities
are simple poles. In this limit the content of unitarity is
that the residues of each such pole in a $N+M$ particle
amplitude must factorize into a product
of scattering amplitudes for $N+1$ and $M+1$ particles. Unitarity further
requires these residues to obey positivity constraints Fig.\ref{factor}. 

This program took the young particle theorists of the time by storm,
whose intense activity led to a rapid development from 5 particle
amplitudes
to $N$ particle amplitudes in a matter of several months. The amplitudes were
discovered by others, but Mandelstam and his colleague Bardakci
\cite{Mandelstam:1969c},
and, independently, Fubini and Veneziano \cite{fubiniveneziano}
were the first to prove the necessary factorization properties
of the $N$ particle amplitude, including a determination of
the degeneracies of the mass levels. More precisely they determined
an upper bound on the degeneracies, because in their
initial factorization some of the factored terms
were not positive in the elastic forward case. If there were
to be no ghosts in the factorization there should be an alternate factorization
with a smaller degeneracy. If no such factorization were possible the
model would have ghosts and therefore violate unitarity.
In their work, Bardakci and Mandelstam found the
first of an infinite string of  Ward-like identities which could
enable this alternate factorization. 

Miguel Virasoro was the one who identified the necessary Ward-like identities
\cite{virasoroelln}.
They were described in the operator formalism developed by Fubini,
Gordon, and Veneziano, in terms of a set of operators $L_n$ which obeyed the 
Virasoro algebra
\bea
{}[L_n,L_m]&=&(n-m)L_{n+m}+\frac{D}{12}(n^3-n)\delta_{n,-m}
\eea
where $D$ is the space time dimension. Provided that the intercept
of the leading Regge trajectory is unity, the operators 
$A_n=L_n-L_0+n-1$ for $n>0$ annihilated physical states. The identity found
by Bardakci and Mandelstam was the $n=1$ case $A_1$ \cite{thorndepend}. 
The requirement of unit
intercept was an early disappointing sign that the known dual resonance
amplitudes could not give a precise description of strong interactions.
Indeed, it predicted a massless vector meson, which was a far cry from
the mass of the $\rho$ meson, the lowest mass spin 1 hadron. On the other
hand the corresponding intercept for the Virasoro dual resonance models
was 2, corresponding to a massless spin 2 particle. For hadrons that was even
worse, but it was a tantalizing early hint that dual models
might have something to say about quantum gravity!

Although Virasoro had found enough Ward-like identities to remove ghosts
from the DRM, it took a while before it was finally proved that's what they did.
\cite{goddardt,brower}. Those proofs made clear that the ghost issue was
highly nontrivial, because their absence required $D\leq26$ for the
generalized Veneziano and Virasoro models. DRM were inconsistent
if the space-time dimension was too high. At first glance this
would not seem to be a serious problem since $D=4$ in our universe.
But difficulties with perturbative unitarity only disappeared if
$D=26$. Some of these dimensions could be compact, but they would still
make the particle spectrum much too rich for our world.

The model just discussed is the bosonic DRM. Despite its freedom
from ghosts for $\alpha_0=1$ and $D=26$, it still possessed a 
flaw: a particle of negative mass squared, a tachyon, when
$\alpha_o(m_o^2)=\alpha^\prime m_o^2+1=0$ in the open (Veneziano) sector and
$\alpha_c(m_c^2)=\alpha^\prime m_c^2/2+2=0$ in the closed (Virasoro) sector.
Sometimes a tachyon in an approximate spectrum signals that
one is perturbing about the wrong ground state, so that the flaw
is not necessarily fatal. But the problem of finding an alternative
ground state has so far evaded solution. 

One can, alternatively,  attribute the problem to the 
absence of spin degrees of freedom in
the model. All of the angular momentum of the massless spin 1 particle is
from orbital motion so there is a ``lighter'' state, the tachyon.
In addition there was no place for spin 1/2 fermions in the bosonic DRM.
The introduction of spin degrees of freedom into DRM thus become a major
focus of activity. Since we were no longer tolerating ghosts of any kind,
Mandelstam's scheme of appending spin factors to the bosonic
amplitudes, as in 
\cite{Mandelstam:1969e,Mandelstam:1970a,Mandelstam:1970b,Mandelstam:1970c}, 
would not do. The way the Dirac particle avoids such ghost states is 
by projecting them out by the Dirac equation itself: $(-i\gamma\cdot\partial
+m)\psi=0$. So there must be an intimate interplay between orbital
and spin degrees of freedom.

There were two major attacks on this problem. To explain them I need to make 
brief mention of the operator formalism introduced by Fubini, Gordon,
and Veneziano \cite{fubinigv}. 
They noticed that the factorization of the open DRM could be
described in terms of harmonic oscillator raising and lowering
operators $a_n^\mu$ with commutation relations $[a_n^\mu,a_m^\mu]=\eta^{\mu\nu}
n\delta_{n,-m}$, where $\eta$ is the mostly positive Minkowski metric 
and $n=0,\pm1,\pm2,\ldots$.
Also $a_0^\mu=\sqrt{2\alpha^\prime}p^\mu$ with $p$ the energy momentum of
the resonance. The mass squared of the resonance is then given by
the operator $\alpha^\prime m^2=-1+\sum_{n=1}^\infty a_{-n}\cdot a_n$, 
and the 
resonance states are in 1-1 correspondence with monomials of the $a_{-n}^\mu$
applied to a ground state $\ket{0,p}$. As Nambu, Nielsen, and Susskind
more or less independently observed, the $a$'s are just 
the normal mode operators of 
a ``string'' $x^\mu(\sigma,\tau)$, which, however, vibrates in spacetime 
rather than space.
Nambu \cite{nambuaction} then postulated the correct dynamics for the 
string which explained how the  ghostly oscillations 
in time $x^0$ were redundant.

But these stringy insights were not directly used in the construction of
DRM with spin degrees of freedom. One approach by Bardakci and Halpern (BH)
\cite{bardakcihalpern}
was to add fermionic oscillators $\psi_r^\alpha$ where $\alpha$ is a Dirac spinor
index. They initially assumed $r=\pm1/2,\pm3/2,\ldots$. The spinor
indices are responsible for additional ghosts. To decouple them
BH postulated Ward-like identities which added terms of the
schematic spin-orbit form ${\bar\psi}\gamma^\mu\psi a_\mu$ to Virasoro
operators. The cubic non-bilinear form of such a term made
it difficult to work with, and progress on the study
of these models  was slow. The other approach. initiated by Ramond
\cite{ramond} 
with important development by Neveu and Schwarz \cite{neveuschwarz}
(RNS) and me, introduced
anticommuting Lorentz vectors $b_n^\mu$, $n=0,\pm1,\ldots$ (R fermion sector) or
$b_r^\mu$, $r=\pm1/2,\pm3/2,\ldots$. (NS boson sector). These new operators
contributed to the Virasoro generators in a bilinear way.
The time components of the $b$'s
introduce more ghosts, but Ramond proposed Ward-like identities
$F_n=0$ where $F_n=\sum_k b_{-k}\cdot a_{k+n}$. These operators being only 
bilinear were much easier to work with than the BH setup. 
For this reason the new DRM
amplitudes were the first to be constructed and analyzed in the
RNS language.

The RNS
amplitudes were shown to be ghost free for $D\leq10$ \cite{goddardt,corrigang}. 
The leading open Regge trajectory
still had unit intercept and the closed Regge trajectory still
had intercept 2. But there were still tachyons, though these were on
lower trajectories than in the bosonic model. 

Interestingly, very
early on (Spring, 1971) Mandelstam remarked to me
that the open tachyon in the mesonic RNS amplitudes, the ``pion'',  
being in the odd G-parity
sector, would decouple if the meson spectrum were restricted to even
G-parity. I mentioned this little gem to my colleagues
at CERN during my year-long tenure there.
Later (1976) Gliozzi, Scherk, and Olive (GSO) observed 
\cite{gso} that
the closed tachyon could also be decoupled if, in addition to the
even G-parity restriction, noticed by Mandelstam, a Weyl restriction were
made on the open fermions. After the even G-parity and GSO projections,
the remaining resonance spectrum fell into multiplets of supersymmetry,
giving birth to the superstring. The superstring amplitudes were
Regge behaved and implied a ghost-free
and tachyon free spectrum that included a massless spin 2 boson. 
I think it is important to note that this construction of superstring
provided theoretical physicists with a fully consistent alternative
paradigm for the description of physical phenomena,
 distinct from quantum field theory. 

In addition to his G-parity insight for the RNS amplitudes,
Mandelstam made substantial contributions to the
BH approach to adding spin to DRM \cite{Mandelstam:1973a,Mandelstam:1973b}.
In the first of these papers, he went some distance to disentangling
the $K$-degeneracy problem in the nonadditive BH models. These models
supported two commuting Virasoro algebras, one of which is chosen to
supply the Ward identities, and the other of which predicted an infinite 
degeneracy of the particle spectrum. Mandelstam showed that the
$K$-degeneracy was a kind of gauge symmetry which can be handled
by imposing a gauge condition. The second paper constructed
a simple viable nonadditive model that seemed to reproduce
many features of the the RNS models, including a critical dimension of
$D=10$. He conjectured but did not
definitively prove that they were in fact the same models. This
result presaged later work by Green and Schwarz \cite{greenschwarz} 
who constructed
a BH-style formulation of the superstring that was manifestly supersymmetric.
Their spinor valued operators, however, were integer moded in
contrast to the BH choice of half integer modes. The half integer
choice would have violated supersymmetry.

\section{Lightcone Interacting String Formalism}
Mandelstam's papers, which develop 
the lightcone interacting string formalism  
(\cite{mandelstambosonic} for the bosonic string and 
\cite{mandelstamnsr}
for the Ramond-Neveu-Schwarz string), were essential for the
future development of string theory. First of all they completed
the program, initiated by Nambu's hypothesis \cite{nambuaction} (see also
\cite{goto})
that the dual resonance model scattering
amplitudes \cite{veneziano} were based on the dynamics
of a relativistic string. The
lightcone quantization of a single noninteracting
relativistic string \cite{goddardgrt} had shown that the
spectrum and degeneracies of excited resonances, implied by the
pole singularities of dual resonance amplitudes, 
coincided (for the critical dimension $D=26$)
with the spectrum and degeneracies of the Nambu-Goto string. 
It was Mandelstam's work that definitively established 
that the scattering amplitudes themselves 
followed from string breaking and joining
processes described by his path integral representation of the
light cone quantum dynamics of the Nambu-Goto string. 
Secondly, by revealing the underlying physics of dual models,
these papers laid the foundation for going beyond string perturbation
theory (notably by summing planar interacting
string diagrams). Here I will try to give a
simplified overview of these papers, accenting these two aspects.
I shall limit the detailed discussion to the bosonic string 
\cite{mandelstambosonic}. I shall only briefly mention his extension of 
the formalism to
the RNS string \cite{mandelstamnsr}, which included, as a tour-de-force, 
the first calculation of amplitudes 
involving four or more external fermions.  

\subsection{Interacting string path integral on the lightcone}
\begin{figure}[ht]
\begin{center}
\includegraphics[width=2.9in,height=1.7in]{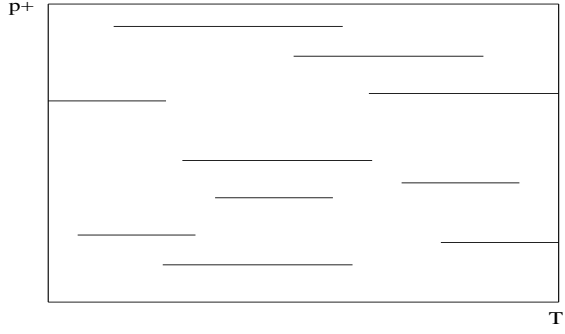}
\caption{Typical interacting string diagram with two incoming strings
three outgoing strings and 7 loops}
\label{typical}
\end{center}
\end{figure}

The operator quantization of the Nambu-Goto action
in lightcone parameterization \cite{goddardgrt} is equivalent 
to path integration over the transverse string coordinates 
${\bfs x}(\sigma,\tau)$
defined on a rectangular domain $0<\sigma<p^+=(p^0+p^z)/\sqrt2$, 
$0<\tau=i(t+z)/\sqrt{2}<T$, with (imaginary time) action given by
\bea
S_{l.c.}=\frac{1}{2}\int_0^Td\tau\int_0^{P^+}d\sigma\left[\left(
\frac{\partial{\bfs x}}{\partial\tau}\right)^2+T_0^2\left(
\frac{\partial{\bfs x}}{\partial\sigma}\right)^2\right]\nonumber
\eea
Then Mandelstam's proposal for a typical contribution to the evolution
of a system of strings initially at $\tau=0$ and finally at $\tau=T$
is simply to alter the geometry of the $(\sigma,\tau)$ 
domain, as shown in Fig.~\ref{typical}.
%\noindent
The horizontal lines in this diagram are internal boundaries
 where open strings end, so the diagram describes 
the time evolution of a system of open strings which
break and rejoin from time to time.

For the open string in the critical dimension, the 
worldsheet path integral uses the lightcone action 
for the free open string. The complete contribution to
the evolution amplitude is obtained by doing the path integral over
${\bfs x}$ summing over all numbers of horizontal lines and
integrating over their lengths and locations in $\sigma,\tau$.

As Mandelstam showed \cite{mandelstambosonic}
the dependence of the diagram on the initial and final
${\bfs x}(\sigma)$ is obtained by solving a suitable
Poisson equation via Neumann functions.
But there is additional dependence on the $P^+_i,\tau_i$,
which characterize the geometry of the diagram,
given by the measure factor, which is in turn related to the determinant
of the Laplacian defined on the domain represented by the
diagram. Since each component of ${\bfs x}$ supplies an identical
determinant factor this measure dependence must be of the
form 
\bea
{\cal M}(P^+_i,\tau_i)&=& [\mu(P^+_i,\tau_i)]^{D-2}
\eea
Thus it is evident that Lorentz invariance can be achieved for
{\it at most} one value of $D$. In his pioneering papers on the subject
\cite{mandelstambosonic},
Mandelstam inferred the measure factor $\mu$ by establishing
first that his formalism was Lorentz covariant for $D=26$, and
then, setting $D=26$, he obtained the measure by evaluating
it in a special Lorentz frame that simplified the calculation.
From the known $D$ dependence, this determines 
the measure $\mu$.
In later work \cite{mandelstamdet} he calculated the 
necessary determinants directly, as I shall briefly describe in the
following. 
\subsection{The 3 string vertex}
\begin{figure}[ht]
\begin{center}
\includegraphics[width=2.3in]{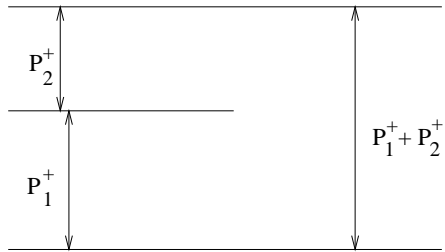}
\caption{Vertex diagram for three open strings}
\label{3vertex}
\end{center}
\end{figure}

I shall explain the importance
of the measure factor in the case of the 3 open bosonic string vertex
represented in Fig.~\ref{3vertex}.
Since the path integral automatically
reproduces the properly normalized probability amplitude,
Lorentz covariance requires an answer of the form: 
\bea{\rm Vertex}&=& \frac{1}{\sqrt{P^+_1P^+_2(P^+_1+P^+_2)}}
F_{h_1h_2h_3}(m_1^2,m_2^2,m_3^2)
\eea
where $h_i$, $m_i$ are the helicities and masses of the
three open string states. The factors of $\sqrt{P^+_i}$ are simply
the lightcone analogs of the familiar
$\sqrt{E_i}$ factors in relativistic probability amplitudes.
They can only come from the measure factor of the path integral,
and prescribe that under the scaling
$P^+_i\to \lambda P^+_i$, the measure factor should scale as
$\lambda^{-3/2}$.

By defining the path integral on a worldsheet lattice, a direct 
calculation shows that the
measure factor scales as $\lambda^{-(D-2)/16}$ for bosonic string
\cite{gilest}, which is compatible with Lorentz invariance only
if $D=26$ the critical dimension. It is
important to appreciate that this scaling behavior
is entirely due to the sharp $360^\circ$ corner in the worldsheet boundary
where the two strings join. This can be inferred from an inspiring paper
by M. Kac \cite{kacdrum}.
Kac analyzed the eigenspectrum of the Laplacian defined on 
a polygonal domain, bounded by straight line segments meeting at various
angles $\theta_i$. He proved the result
\bea
\Tr\ e^{t\nabla^2/2}\sim \frac{{\rm Area}}{2\pi t}-\frac{{\rm  Perimeter}}{
4\sqrt{2\pi t}}+\sum_{\rm corners}\frac{1}{24}\left(\frac{\pi}{\theta_i}
-\frac{\theta_i}{\pi}\right)+o(1)\nonumber
\eea
The lightcone vertex  is a $360^\circ$  corner, so putting 
$\theta=2\pi$ gives
\bea
\frac{1}{24}\left(\frac{\pi}{\theta}-\frac{\theta}{\pi}\right)\to-\frac{1}{16}
\eea
which agrees with the lightcone lattice result.  
McKean and Singer \cite{mckeans}
generalized Kac's result to an arbitrary smooth geometry. 

In \cite{mandelstamdet}, Mandelstam evaluated the determinants that
enter the measure factor for his path integrals by exploiting
the conformal transformation properties of the determinant of
the Laplacian:
\bea
-\frac{1}{2}\delta(\ln(-\nabla^2))&=&\frac{1}{24\pi}
\int dA g^{ab}\frac{d\sigma}{dz_a}\frac{d\sigma}{dz_b}
+\frac{1}{12\pi}\int d\ell k\sigma
+\frac{1}{24\pi}\int dA R\sigma\nonumber\\
&&+\frac{1}{24}\sum_{\rm corners}\left(\frac{\pi}{\theta_i}-
\frac{\theta_i}{\pi}\right)\sigma(z_i)
\eea
where the new metric is $e^{2\sigma}$ times the old metric.
The terms involving the curvature $k, R$ and corners refer to the
{\it old metric}, which Mandelstam took to be the image
of the interacting string diagram in the upper half (whole)
$z$-plane for open strings (closed strings). The determinant
of the Laplacian in the $z$-plane as a function of the radii
of the semi-circles (circles) can be easily inferred in the limit
where all but one of these radii are small and the last one is 
large.

The boundaries
of the lightcone diagram at large initial and final times
map to semi-circles (circles) in the upper half (whole) $z$-
plane. The only corners in the $z$-plane are the $90^\circ$
corners where the open string semi-circles meet the real
axis. For closed string amplitudes the corner terms are
absent all together. Mandelstam's evaluation neglected 
corner terms. However it correctly gave the vital $P^+$ dependence associated
with the $360^\circ$ corner at the interaction point in the
lightcone diagram, because that corner is the
image under the conformal transformation of a nonsingular point 
on the real axis of $z$-plane: The conformal transformation
formula, with corner terms neglected, correctly reproduces Kac's result 
for those corners in 
the target manifold that are images of nonsingular points
in the original manifold! 

The neglect of the contributions from the $90^\circ$ corners,
which are present in both the target and original manifolds, is
inconsequential for the calculation of the $S$-matrix.
This is because those same corner contributions,
which are multiplicative, are present
in the free string propagator, which determines the
normalization of the asymptotic states of the $S$-matrix.
Even if they had been properly included, they would have nonetheless
completely cancelled out of the $S$-matrix! The only corner that
matters, the $360^\circ$ one at the interaction point, is
correctly treated in Mandelstam's evaluation.
\subsection{Scattering Amplitudes for the Superstring}
Before Mandelstam's application of his lightcone interacting
string formalism to the RNS dual resonance models, there
were already formulas for the scattering amplitudes involving
any number of bosonic particles with zero or 2 fermionic particles.
These amplitudes had been obtained using operator techniques
rather than path integrals. The extension of these operator methods
to amplitudes with more than two fermionic particles, however, 
was bogged down.

In this setting Mandelstam's lightcone path integral calculation of 
amplitudes with arbitrary numbers of fermions \cite{mandelstamnsr}
had a stunning impact. The measure part of these calculations
required the calculation of the determinant of
the worldsheet Dirac operator rather than the worldsheet Laplacian.
However, with the RNS boundary conditions the $z$-plane images
of the joining point are nonsingular points, so the important
part of the measure can be inferred again by conformal mapping,
just as in the bosonic case. The boundary conditions
on either leg of the 360$^\circ$ angle at the joining point 
are $S_1=S_2$ on one side
of the angle vertex but $S_1=-S_2$ on the other side.
Then because the worldsheet fermions have conformal weight 1/2
the conformal transformation to the upper half plane which
rotates the line on which $S_1=-S_2$ by $180^\circ$ flips that
line's boundary condition to $S_1=S_2$, so there is no discontinuity
at the image of the interaction point. This enabled Mandelstam to
write down the amplitude for any number of fermions effortlessly.

Applying the lightcone path integral methods to the 
Green-Schwarz lightcone
formulation of the superstring is not so simple
\cite{Mandelstam:1985c,mandelstamdet}. This is because
the boundary conditions in this case are $S_1=S_2$ for {\it all} boundaries
(because the worldsheet fermion fields are integer moded on all external
legs of a vertex function). Thus the image of the joining point
on the upper half plane is the location of a discontinuity in the
boundary conditions on the real axis of the upper half plane. This is
a $180^\circ$ ``corner'' that must be included in the McKean-Singer
conformal transformation formula.
Thus the conformal transformation is not the
only source of $P^+$ factors: the $z$-plane determinant is no
longer trivial. Since the superstring amplitudes can also be obtained
in the RNS formulation it was possible to figure out the 
required $z$-plane determinant indirectly. It would 
certainly be interesting to fill this gap with a direct calculation, 
for example, along the lines used for Dirichlet-Neumann $180^\circ$ corners in 
\cite{thorndet}.
\section{Quark Confinement}
One of the frustrations of string theory was that, although it
was discovered in the effort to understand the strong interactions,
it fell short of reaching that goal. The string mass spectrum, 
inevitably including massless vector and massless tensor particles,
was more suited to gauge and gravity theories! After the
discovery of asymptotic freedom and charm in 1973 and 1974,
overwhelming evidence accumulated
that a nonabelian gauge theory of quarks
and gluons based on color SU(3) gauge group was the correct 
theory of strong interactions. For this choice to be valid, however,
the hypothesis of quark confinement, that quarks and gluons are
permanently trapped inside of hadrons must hold. Understanding 
the theoretical physics underlying quark confinement attracted
Mandelstam's attention and dominated his work for the rest of
the decade.
\subsection{Early papers on gauge theory and gravity}
Before describing Mandelstam's work on quark confinement, I would
like to describe briefly four interesting papers he wrote on the
fundamental formulation of gauge theories and quantum gravity
\cite{Mandelstam:1962a,Mandelstam:1962b,Mandelstam:1968a,Mandelstam:1968b}.
Gauge theories and general relativity in their standard treatment
use dynamical variables which are not directly observable: the
vector potential $A^\mu$ in the first case and coordinate systems and
metric $g_{\mu\nu}$ in the second. All four papers have the ambitious goal
of setting up the fundamental quantum mechanics of such systems without
the initial introduction of such quantities. The case of quantum 
electrodynamics is easiest to explain. The electromagnetic 
fields $F_{\mu\nu}$ have
direct physical meaning and are included in the list of fundamental
quantities. 

In the standard formulation a charged field satisfies equations
in which the vector potential, related to the fields by $F_{\mu\nu}
=\partial_\mu A_\nu-\partial_\nu A_\mu$, makes a direct appearance
$-(\partial-iQA)^2\phi+m^2\phi=0$. Then quantization proceeds by
promoting $\phi, A$ to operators subject to commutation relations.
In \cite{Mandelstam:1962a}, Mandelstam instead promotes $F_{\mu\nu}$
and a path dependent charged field $\phi(P_x)$ to operators. In
the formulation with potentials this path dependent field would
be
\bea
\phi(P_x)&=&\phi(x)\exp\left\{iQ\int_{P_x}dx^\mu A_\mu(x)\right\}
\eea
Mandelstam replaces such a relation with a rule giving the path
dependence
\bea
\delta\phi(P_x)&=&iQ\delta\sigma^{\mu\nu}F_{\mu\nu}\phi(P_x)
\eea
where $\delta\sigma^{\mu\nu}$ is a surface element spanning the closed
curve formed by $\delta P$ and the original path. The homogeneous
Maxwell equations guarantee that a global deformation of the path
is independent of the surface chosen to span the change.

Clearly quantizing path dependent dynamical variables involves
many complicating issues, and the papers which confront and
deal with them are vintage Mandelstam. He systematically identifies
each complication and through thorough and close reasoning shows
how to handle it, The second paper \cite{Mandelstam:1962b} applies
the same principles to the quantization of Einstein's theory
of gravity. The fundamental variables are the curvature tensor and
matter fields, {\it both} of which are given path dependence.
The setup is very much more involved, but Mandelstam forthrightly
deals with every obstacle in his path.

The two 1962 papers are mostly concerned with the fundamental
setup of the new formalism without regard to the strength of interactions.
The 1968 papers \cite{Mandelstam:1968a,Mandelstam:1968b} are concerned
with obtaining the Feynman rules of perturbation theory from
Mandelstam's gauge or coordinate independent formalism. in gauge 
theories and gravity respectively. In addition to electrodynamics,
he also obtains the Feynman rules for nonabelian gauge theory.
His results are in accord with those of the standard procedures,
including the rules for Feynman-Fadeev-Popov ghosts. 

These four papers express a vision of quantum gauge theories and quantum gravity
which is in principle more in tune with their classical analogues.
A modern implementation of this vision can be found in Wilson's formulation
of lattice gauge theory, in which the fundamental dynamical variables
are group elements $U_L$ (as opposed to Lie algebra elements) assigned to
the links $L$ of the lattice. Instead of the $-F^2$ term in the Lagrangian,
the lattice theory uses the product of $U's$ about each plaquette.
Path dependent observables such as Wilson loops play a central
role in extracting physical information from lattice gauge theory
\subsection{Understanding QCD}
In the mid to late 70's Mandelstam's research concentrated on
investigating the physics of quark confinement in QCD. By that time
Nielsen and Olesen \cite{nielseno}, motivated by the close connection between
superconductivity and the Higgs mechanism in abelian gauge theory,
had found stable classical solutions of the coupled gauge and 
Higgs equations of motion in which a long magnetic flux tube 
centered on a line was stabilized by its interaction with
the Higgs field. In the semi-classical approximation these
classical solutions can be interpreted as new quantum states beyond the 
vacuum and gauge and Higgs particles.

In a superconductor such states exist 
because of the Meissner effect, in which the magnetic field
${\bfs B}$ is expelled from the interior of a supeconducting medium.
If north and south magnetic monopoles were placed at finite separation
in a superconductor that filled all of space, the magnetic flux from
one would have to end on the other. But the Meissner effect would require
the flux to be collimated in a tube sufficiently thin, so that the magnetic
field would have the strength to destroy superconductivity inside
the flux tube. This would cost an energy proportional to the separation
$R$ of the monopoles, i.e. there would be a linear confining potential
energy between the monopoles. Nielsen and Olesen and also Nambu observed that
if quarks were monopoles, this would provide a mechanism for quark
confinement.

The reason why the vacuum of the Higgs mechanism is a superconductor
can be understood in terms of the vacuum polarization tensor
$R^{\mu\nu}=(q^\mu q^\nu-q^2\eta^{\mu\nu})R(q^2)$. The $q$ dependent
dielectric parameter of the vacuum, thought of as a medium, 
is related to $R$ by the
formula $\epsilon(q^2)=1+R(q^2)$, with the static dielectric constant
the limit $q\to0$.. The potential of the
Higgs field is in the shape of a Mexican hat so that the phase
of the Higgs field is a massless scalar, which produces a massless
pole in $R(q^2)\sim K/q^2$. Thus $\epsilon(q^2)\to\infty$ as $q\to0$.
But in a Lorentz covariant gauge theory, $\epsilon(q^2)\mu(q^2)=1$
with $\mu$ the magnetic permeability.
It follows that $\mu(0)=0$, so  the Higgs vacuum is a perfect diamagnet,
and so exhibits the Meissner effect.

In \cite{Mandelstam:1974d,Mandelstam:1974c}, Mandelstam studied this effect
in the SU(2) nonabelian gauge theory. There is of course the Dirac
quantization condition that the monopole strength must be integer multiples
of the reciprocal of nonabelian charge. He showed that only monopoles of unit
strength would produce a flux tube. (For SU(n) flux tubes would form between
monopoles of strength $1,2,\ldots n-1$.) The flux is defined modulo $n$, and
he showed for $n=2$ that a two unit flux tube solution could be continuously
deformed to the vacuum. As a model of confinement, the quantization condition
on the monopole charge would conflict with asymptotic freedom, a
demonstrable property of QCD: the magnetic charge would have to 
blow up at short distances.

Moreover asymptotic freedom for QCD implies that the dielectric parameter
$\epsilon(q^2)$ increases as $q^2$ increases. To see this recall that the
effective charge is $g^2(q^2)=g_0^2/(4\pi\epsilon(q^2))$. Asymptotic freedom
means that $g^2(q^2)$ decreases with $q^2$, implying that $\epsilon(q^2)$
increases with $q^2$. or
$\epsilon$ decreases as $q^2$ decreases. The static properties
of the medium are given by the limit $q\to0$. This
is {\it not} indicative of the behavior of a diamagnetic medium--it 
is rather the behavior
of a paramagetic medium, since $\mu=1/\epsilon$ increases as $q^2$ decreases.
Quark confinement in QCD should correspond to a perfect dia-electric
and, by Lorentz invariance, a perfect paramagnetic medium. 
So Mandelstam proposed replacing
the Higgs field by a field which creates a magnetic
monopole. He then proposed a model of the QCD vacuum that contained a
condensate of this field. In other words his proposal is essentially the
dual of a superconductor in which electric and magnetic quantities are
interchanged. The question of whether such a vacuum ansatz minimized
the energy was left unresolved in these papers.

\subsubsection{Electric and magnetic variables in gauge theory.}
In his next paper on this subject \cite{Mandelstam:1978b} Mandelstam embarks
on a thorough and systematic study of electric/magnetic duality
in nonabelian gauge theories. One has the usual magnetic
vector potentials to start with, but Mandelstam constructed
electric vector potentials in terms of them. In addition he considered operators
that create loops of electric or loops of magnetic flux. The former
are simply the standard Wilson loops, and the latter are related to 
loops considered previously by 't Hooft. He then used these constructions
to delineate the possible phases of the gauge theory. Two of the possible phases
are magnetic confinement a la Nielsen and Olesen and electric confinement
a la Wilson. It was hoped that QCD picks the latter phase.
Neither of these phases would support massless particles.
A third phase with massless gauge particles could not be ruled out. 

In four spacetime dimensions, the explicit construction of functions of
quantum field operators is fraught with the distracting presence of
ultraviolet divergences. This clouds the details of the
attempted construction of ``electric'' variables in terms of
``magnetic'' variables. Thus it is reassuring that Mandelstam was able to
implement rigorously an explicit operator mapping 
of the quantum field operators in 
two popular toy models in two spacetime dimensions. In 
\cite{Mandelstam:1975a,Mandelstam:1976a} he demonstrated the
equivalence of the massive Thirring model and the sine-Gordon
scalar field model, both in two spacetime dimensions. The first
model is a QFT of a Fermi field $\psi$ with
an interaction term quartic in the fields. In the massless case
it had been exactly solved by Thirring. The second toy model is 
a QFT of a scalar field $\phi$ with an interaction potential
$V(\phi)=(1-\cos\phi)$. It was interesting because it admitted
soliton solutions, which were exact static classical  solutions 
with the property
$\phi(x=\infty)=\phi(x=-\infty)+2\pi$. Evidence for the equivalence
of the two models had already been given by Coleman. But Mandelstam
was able to construct the Thirring field $\psi$ as an explicit
operator function of the sine-Gordon field $\phi$. Since the
model is two dimensional, he had complete control over ultraviolet
divergences. An interesting feature of Mandelstam's mapping 
between the two theories is that the weak coupling regime of one
maps onto the strong coupling regime of the other, and vice versa.

\subsubsection{An approximation scheme for QCD}
If all quarks are massless, there are no free parameters in QCD
with the gauge group SU(3):
It is classically scale invariant, but ultraviolet divergences
break scale invariance so that the apparently free coupling
parameter is exchanged for a single mass scale $\Lambda$.
Alternatively the coupling depends on the momentum $\alpha_s(q^2)$,
and $\Lambda$ specifies the scale at which $\alpha_s$ has some standard
value, e.g. $\alpha_s(\Lambda^2)=1$. The physical analogy associating 
confinement with 
a dual Meissner effect is insightful, and Mandelstam's effort to
implement it via monopole condensation was convincing in a qualitative
way..
But in the inspiring paper \cite{Mandelstam:1979a}, he takes a different tack.
Instead of trying to construct an explicit model for the 
monopole condensate, he argues that it is enough if magnetic effects
are sufficiently enhanced in the gluon propagator at low frequencies 
and wave numbers. By
replacing the sum over Feynman diagrams with the set of
Schwinger-Dyson equations for multigluon amplitudes, one can explore
the effect of such enhancements on the physics of confinement.

For instance, a static confining linear potential in a
Coulomb gauge would arise
from the Fourier transform of a small $q$ behavior $({\bfs q}^2)^{-2}$. In the
context of the gluon propagator in a Lorentz covariant gauge this
behavior would translate to $(q^2)^{-2}$. Mandelstam's strategy was
to explore the consistency of such behavior with the Schwinger-Dyson
equations, and if so to find (or at least prove the existence of)
a solution that exhibits it. The full set of Schwinger-Dyson equations
was too complicated, so Mandelstam truncated them to get a tractable
equation. He was able to establish that the desired small $q$ behavior
was consistent with the truncated equations Through a rough iterative numerical
solution of the equations he obtained pretty convincing evidence that
a suitable solution existed. Although the truncated S-D equations
could not be achieved as the limit of a parameter of the theory
(there are none!), it was clear what graphs had been omitted
in the approximation. At least in principle, one could hope to restore
them gradually to improve the solution. In the rest of the paper,
however, he stuck with his original truncation together with
well-described approximations for dealing with multigluon amplitudes
and quark propagators. He closed with a discussion of a Bethe-Salpeter
equation for the propagation of a quark antiquark system using
the ingredients just described. One output 
was linearly rising Regge trajectories. This was not surprising given the
infrared behavior of the gluon propagator.
\section{Closing Tribute}
Stanley Mandelstam was a truly impressive theoretical physicist. 
His work has touched on nearly all of the significant
problems in quantum field theory and elementary particles of
his lifetime. Each of his published articles is a significant and
inspiring achievement and a gold mine of closely reasoned theoretical
physics. But Stanley's impact extended way beyond his published
work. I am especially aware of his impact on the development of
string theory, since that was a central focus of my early research.
The discovery and development of dual resonance models, a.k.a 
string theory, was a spectacular shift from the existing theoretical
paradigm of quantum field theory. Young particle theorists, who
participated in these developments, were privileged
to experience the thrill and
excitement of creating something truly new, even though the
new paradigm fell short of its original goal. But their
work might never have been pursued without the support and
encouragement of three masters in the older generation: 
Stanley Mandelstam, Yoichiro
Nambu, and Sergio Fubini.  Each, in his own way, made pioneering advances
that enabled and guided the developments. These three men
not only shared their considerable achievements
and wisdom with us youngsters, but they
also came down into the trenches and shared a good deal of the
burden with us. As one of Stanley's Ph.D. students, I was 
particularly aware of the ways Stanley would inspire others to
take an important step--always with a light and gentle touch,
often with a one line provocative remark. He set me off on my first
research project in just this way. I don't remember his
exact words, but it was something like: ``this question may well be
soon resolved by others, but you might like to try to understand the 
(Bardakci-Mandelstam) Ward-identity in the operator formalism.''
It was a good problem. Stanley was the go-to person whenever a student
or postdoc at Berkeley reached an impasse. He would usually know
the resolution, but if not he would almost always suggest a
new path which led to success. Every string theorist, young and old,
owes a tremendous debt to Mandelstam, Nambu, and Fubini. They are
all gone now, but their vision and spirit live on.

\vskip14pt
\noindent\underline{Acknowledgments}: I would like to thank Lars Brink
for encouraging me to write this essay on the work of my Ph.D.
supervisor. I also thank Korkut Bardakci for helpful comments,
especially for drawing my attention to \cite{Mandelstam:1960d}.

\newpage
\renewcommand\refname{Articles by Stanley Mandelstam}

\renewcommand\refname{Other References}

\end{document}